\documentclass[twocolumn,aps,superscriptaddress]{revtex4}

\usepackage{amssymb}
\usepackage{amsmath}
\usepackage{graphicx}
\usepackage[normalem]{ulem}
\usepackage[dvips]{color}
\usepackage{appendix}

\begin{document}

\title{Nucleus giant resonances from an improved isospin-dependent Boltzmann-Uehling-Uhlenbeck transport approach}
\author{Jun Xu\footnote{xujun@zjlab.org.cn}}
\affiliation{Shanghai Advanced Research Institute, Chinese Academy of Sciences,
Shanghai 201210, China}
\affiliation{Shanghai Institute of Applied Physics, Chinese Academy
of Sciences, Shanghai 201800, China}
\author{Wen-Tao Qin}
\affiliation{Shanghai Institute of Applied Physics, Chinese Academy
of Sciences, Shanghai 201800, China}
\affiliation{University of Chinese Academy of Sciences, Beijing 100049, China}
\date{\today}

\begin{abstract}
We have studied the isoscalar giant quadruple resonance (ISGQR) and the isovector giant dipole resonance (IVGDR) in $^{208}$Pb based on an improved isospin-dependent Boltzmann-Uehling-Uhlenbeck transport approach using an improved isospin- and momentum-dependent interaction. With the isoscalar nucleon effective mass and the nucleon-nucleon cross section which reproduces respectively the excitation energy and the width of the ISGQR strength function, the slope parameter of the symmetry energy and the neutron-proton effective mass splitting are constrained respectively within $36<L<62$ MeV and $0.08\delta<(m_{n0}^*-m_{p0}^*)/m<0.42\delta$, by comparing the resulting centroid energy of the IVGDR and the electric dipole polarizability with the experimental data. It is found that nucleon-nucleon collisions have considerable effects on the resulting electric dipole polarizability, which needs to be measured more accurately in order to pin down isovector nuclear interactions.
\end{abstract}

\maketitle

\section{INTRODUCTION}
\label{introductionhan}

Understanding the microscopic nuclear interaction as well as the nuclear matter equation of state (EOS) is one of the main goals of nuclear physics. Thanks to the great efforts made by pioneer nuclear physicists, so far the uncertainties mainly exist in the isospin-dependent part of the EOS, i.e., the nuclear symmetry energy $E_{sym}$, whose density dependence is generally characterized by the slope parameter $L$ around the saturation density. In the microscopic level, the exchange contribution of the finite-range part of the effective nuclear interaction leads to the momentum-dependent nuclear potential, which is related to the nuclear matter EOS. The nucleon effective mass characterizing the momentum dependence of the nuclear potential can be different for neutrons and protons in the isospin asymmetric nuclear matter. The isospin splitting of the neutron and proton effective mass $m_n^*-m_p^*$ is also related to the symmetry energy through the Hugenholtz-Van Hove theorem~\cite{XuC10,BAL13}. Both the symmetry energy and the neutron-proton effective mass splitting have important ramifications in nuclear astrophysics, nuclear reactions induced by neutron-rich nuclei, and nuclear structures. Reviews on the symmetry energy can be found in Refs.~\cite{Bar05,Ste05,Lat07,Li08,EPJA}, and a recent review on the neutron-proton effective mass splitting can be found in Ref.~\cite{Li18}.

Observables of finite nuclei are important probes of nuclear interactions in nuclear medium at subsaturation densities. Both the isoscalar and isovector excitations of finite nuclei are good probes for the corresponding channels of nuclear interactions and EOSs (see, e.g., Ref.~\cite{Bar19}). The pygmy dipole resonance (PDR) and the IVGDR are typical isovector excitations in nuclei and good probes of isovector nuclear interactions. The former represents the oscillation of the neutron skin against the nucleus inert core, while the later is an oscillation mode in which neutrons
and protons move collectively relative to each other. The strength function of the PDR generally peaks at lower excitation energies compared to that of the IVGDR~\cite{Bar12,Urb12}, while both are sensitive to the symmetry energy which prevents the center-of-masses of neutrons and protons from being away from each other. Typically, various studies have shown that the centroid energy and the electric dipole polarizability extracted from the strength function of the IVGDR are found to be good probes of the symmetry energy~\cite{Tri08,Rei10,Pie12,Vre12,Roc13b,Col14,Roc15,zhangzhen15,zhenghua16}. On the other hand, it is intuitively expected that the frequency of the collective oscillation is sensitive to not only the bulk energy but also to the microscopic nuclear interaction characterized by the nucleon effective mass. Fortunately, the isoscalar nucleon effective mass can be extracted from the excitation energy of the ISGQR~\cite{Boh75,Boh79,Bla80,Klu09,Roc13a,zhangzhen16,Kon17,Bon18}, with the help of the available experimental results from $\alpha$-nucleus scatterings~\cite{ISGQRex1,ISGQRex2,Van15}. For a given isoscalar nucleon effective mass, more recent studies have shown that the centroid energy and the electric dipole polarizability can be used to extract the nuclear symmetry energy and the neutron-proton effective mass splitting simultaneously~\cite{zhangzhen16,Kon17}.

Nuclei giant resonances can be studied by both the random phase approximation (RPA) method and transport approaches. Despite the succusses of the RPA method, the width of the strength function is generally missing, unless higher-order contributions~\cite{Ber83}, such as the particle-vibration coupling~\cite{Col94,Lit07}, are taken into account. The Boltzmann transport approach, which has previously been used to extract the EOS and symmetry energy at both subsaturation and suprasaturation densities from heavy-ion collisions (see, e.g., Refs.~\cite{Ver14,Xu19,Col20}), is based on the Boltzmann equation, with the collision term effectively containing higher-order contributions when derived from the von Neumann equation with the n-body density matrix~\cite{Cas90,Aic91}. The collision term leads to the damping of the collective excitation, or equivalently, the width of the strength function~\cite{Bur88,Gai10}. Reproducing correctly the width can be important in obtaining accurately observables related to the moments of the strength function.

In the present work, we study giant resonances in $^{208}$Pb using an improved isospin-dependent Boltzmann-Uehling-Uhlenbeck (IBUU) transport approach. An improved momentum-dependent interaction (ImMDI) is used in the transport approach based on the lattice Hamiltonian framework. Ground-state initializations are achieved with different parameters used in the ImMDI model, and collisions are also improved with the more rigourous energy conservation condition and better Pauli blockings. These theoretical details together with formulas related to nuclei giant resonances are discussed in Sec.~\ref{theory}. We first reproduce both the excitation energy of the ISGQR and its width for $^{208}$Pb by using a proper isoscalar nucleon effective mass and a constant isotropic cross section. The slope parameter of the symmetry energy and the neutron-proton effective mass splitting are then extracted from the centroid energy of the IVGDR and the electric dipole polarizability for $^{208}$Pb. These results are discussed in Sec.~\ref{results}, and a summary is given in Sec.~\ref{summary}.

\section{Theoretical framework}
\label{theory}

\subsection{Effective nuclear interactions}

The potential energy density of the ImMDI model, which can be obtained from an effective two-body interaction with a zero-range density-dependent term and a finite-range Yukawa-type term based on the Hartree-Fock calculation~\cite{Xu10}, has the following form in the asymmetric nuclear matter with isospin asymmetry $\delta$ and nucleon number density $\rho$~\cite{Das03,Xu15}
\begin{eqnarray}
V^{\rm ImMDI}(\rho ,\delta ) &=&\frac{A_{u}\rho _{n}\rho _{p}}{\rho _{0}}+\frac{A_{l}}{%
 	2\rho _{0}}(\rho _{n}^{2}+\rho _{p}^{2})+\frac{B}{\sigma+1}\frac{\rho^{\sigma +1}}{\rho _{0}^{\sigma }}  \notag \\
 & &\times (1-x\delta ^{2})+\frac{1}{\rho _{0}}\sum_{\tau ,\tau^{\prime}}C_{\tau ,\tau ^{\prime }}  \notag \\
 & &\times \int \int d^{3}pd^{3}p^{\prime }\frac{f_{\tau }(\vec{r}, \vec{p}%
 	)f_{\tau ^{\prime }}(\vec{r}, \vec{p}^{\prime })}{1+(\vec{p}-\vec{p}^{\prime})^{2}/\Lambda ^{2}}. \label{MDIV}
\end{eqnarray}%
In the above, $\rho_n$ and $\rho_p$ are number densities of neutrons and protons, respectively, $\rho _{0}$ is the saturation density, $\delta =(\rho _{n}-\rho _{p})/\rho$ is the isospin asymmetry, and $f_{\tau }(\vec{r}, \vec{p})$ is the phase-space distribution function, with $\tau=1(-1)$ for neutrons (protons) being the isospin index. The single-particle mean-field potential for a nucleon with momentum $\vec{p}$ and isospin $\tau$ in the asymmetric nuclear matter with isospin asymmetry $\delta$ and nucleon number density $\rho$ can be obtained from Eq.~(\ref{MDIV}) through the variational principle as
 \begin{eqnarray}
 U^{\rm ImMDI}_\tau(\rho ,\delta ,\vec{p}) &=&A_{u}\frac{\rho _{-\tau }}{\rho _{0}}%
 +A_{l}\frac{\rho _{\tau }}{\rho _{0}} +B\left(\frac{\rho }{\rho _{0}}\right)^{\sigma }(1-x\delta ^{2}) \notag \\
 & & -4\tau x\frac{B}{%
 	\sigma +1}\frac{\rho ^{\sigma -1}}{\rho _{0}^{\sigma }}\delta \rho
 _{-\tau }
 \notag \\
 & & +\frac{2C_{\tau,\tau}}{\rho _{0}}\int d^{3}p^{\prime }\frac{f_{\tau }(%
 	\vec{r}, \vec{p}^{\prime })}{1+(\vec{p}-\vec{p}^{\prime })^{2}/\Lambda ^{2}}
 \notag \\
 & & +\frac{2C_{\tau,-\tau}}{\rho _{0}}\int d^{3}p^{\prime }\frac{f_{-\tau }(%
 	\vec{r}, \vec{p}^{\prime })}{1+(\vec{p}-\vec{p}^{\prime })^{2}/\Lambda ^{2}},
 \label{MDIU}
 \end{eqnarray}%
where the four parameters $A_{u}$, $A _{l}$, $C_{\tau,\tau}$, and $C_{\tau,-\tau}$ can be expressed as~\cite{Xu15}
\begin{eqnarray}
 A_{l}(x,y)&=&A_{0} + y + x\frac{2B}{\sigma +1},   \label{AlImMDI}\\
 A_{u}(x,y)&=&A_{0} - y - x\frac{2B}{\sigma +1},   \label{AuImMDI}\\
 C_{\tau,\tau}(y)&=&C_{l0} - \frac{2yp^2_{f0}}{\Lambda^2\ln [(4 p^2_{f0} + \Lambda^2)/\Lambda^2]},   \label{ClImMDI}\\
 C_{\tau,-\tau}(y)&=&C_{u0} + \frac{2yp^2_{f0}}{\Lambda^2\ln[(4 p^2_{f0} + \Lambda^2)/\Lambda^2]}.   \label{CuImMDI}
\end{eqnarray}
In the above, $p_{f0}=\hbar(3\pi^{2}\rho_0/2)^{1/3}$ is the nucleon Fermi momentum in the symmetric nuclear matter at the saturation density. The isovector parameters $x$ and $y$ are introduced to mimic the density dependence of the symmetry energy, i.e., the slope parameter $L=3\rho _{0}(d E_{sym} /d \rho) _{\rho =\rho _{0}}$, and the momentum dependence of the symmetry potential or the neutron-proton effective mass splitting. The values of the parameters $A_{0}$, $C_{u0}$, $C_{l0}$, $B$, $\sigma$, and $\Lambda$ are adjusted to reproduce the empirical nuclear matter properties, i.e., the saturation density $\rho_{0}$, the binding energy $E_{0}(\rho_{0})$ at the saturation density, the incompressibility $K_{0}$, the symmetry energy $E_{sym}(\rho_{0})$ at the saturation density, the isoscalar potential $U_{0,\infty}$ at the saturation density and at infinitely large momentum, and the isoscalar nucleon effective mass $m^{*}_{s0}$ at the saturation density and at the Fermi momentum. The non-relativistic k-mass in the present study is defined as
\begin{eqnarray}
\frac{m_{n(p)}^{*}}{m}=\left( 1+\frac{m}{p}\frac{\partial U_{n\left ( p \right )}}{\partial p}\right) ^{-1},
\end{eqnarray}
where $m$ is the bare nucleon mass. The isoscalar nucleon effective mass is the same as the neutron or the proton effective mass in the symmetric nuclear matter, while the neutron-proton effective mass splitting in the isospin asymmetric nuclear matter with isospin asymmetry $\delta$ is related to the isoscalar ($m^{*}_{s}$) and isovector ($m^{*}_{v}$) nucleon effective mass through the following relation to the first-order of $\delta$ expansion
\begin{eqnarray}\label{mnp}
m_n^*-m_p^* \approx \frac{2m_s^*}{m_v^*}(m_s^*-m_v^*)\delta.
\end{eqnarray}
Note that $m^{*}_{s}$ and $m^{*}_{v}$ generally depend on both the nucleon momentum and the density of the nuclear matter, but are usually represented by their values at the saturation density and at the Fermi momentum, indicated as $m^{*}_{s0}$ and $m^{*}_{v0}$ in the present manuscript.

\begin{figure}[ht]
	\includegraphics[scale=0.3]{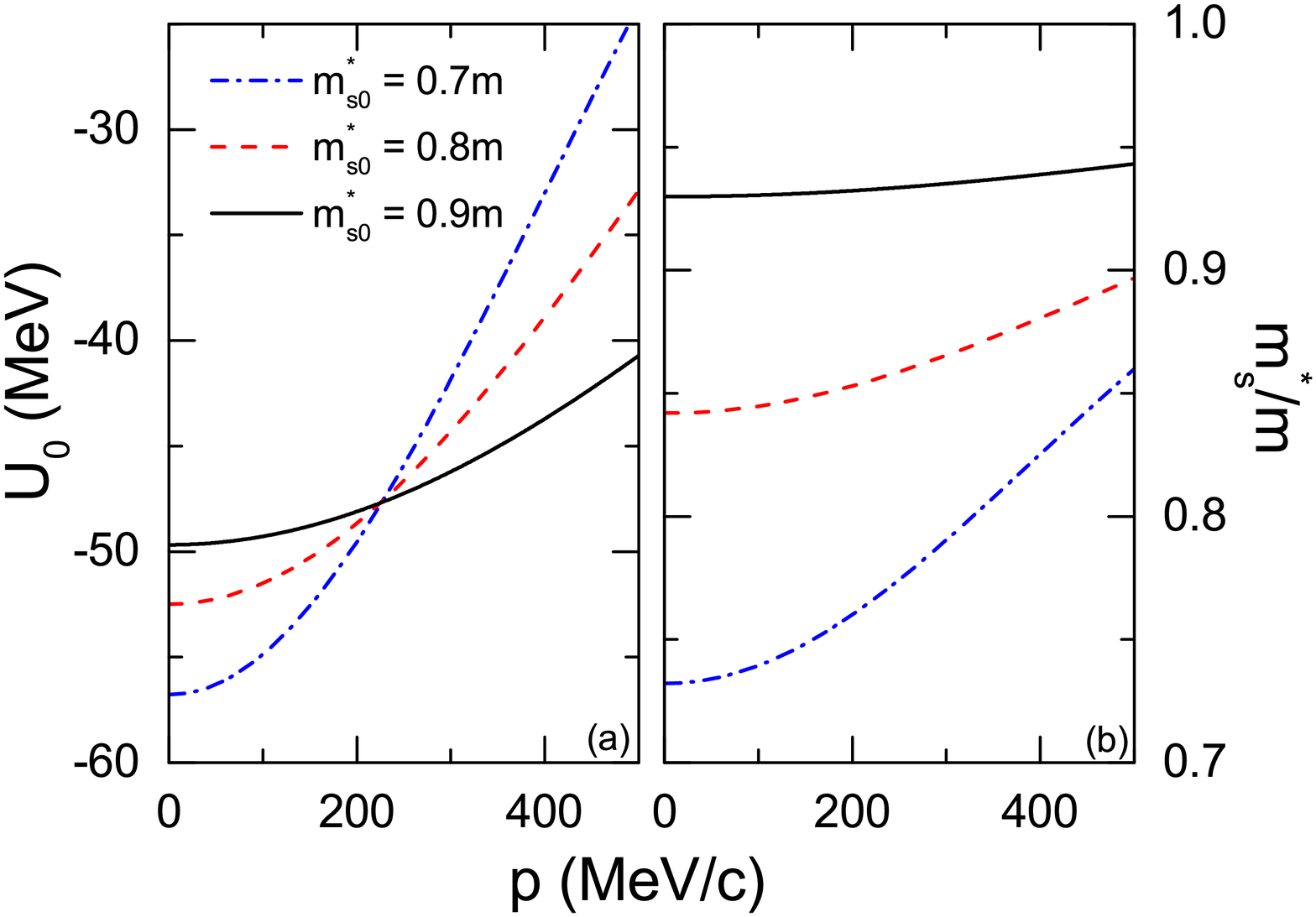}
	\caption{(Color online) Momentum dependence of the isoscalar potential (a) and the isoscalar nucleon effective mass (b) in the nuclear matter at $\rho=0.1$ fm$^{-3}$.} \label{fig1}
\end{figure}

Figure~\ref{fig1} displays the isoscalar potential and the isoscalar nucleon effective mass as a function of the nucleon momentum in the nuclear matter at $\rho=0.1$ fm$^{-3}$, i.e., the average density of a nucleus. All the values of the parameters $A_{0}$, $C_{u0}$, $C_{l0}$, $B$, $\sigma$, and $\Lambda$ need to be adjusted, in order to get different $m_{s0}^*$ but the same $\rho_{0}$, $E_{0}(\rho_{0})$, $K_{0}$, $E_{sym}(\rho_{0})$, and $U_{0,\infty}$, as listed in Table~\ref{T1}. The isoscalar potential is larger (smaller) below (above) the Fermi momentum (about 225 MeV at $\rho=0.1$ fm$^{-3}$) for a larger $m_{s0}^*$, while it is the same at the Fermi momentum for different $m_{s0}^*$ by the model construction. Since the potential below the Fermi momentum is expected to dominate the dynamics of nuclei resonances, a larger $m_{s0}^*$ gives an overall less attractive potential. The isoscalar nucleon effective mass generally increases with increasing nucleon momentum, and its value in the nuclear matter at subsaturation densities is larger than $m_{s0}^*$.

\begin{table}\small
  \caption{Values of parameters and some physics quantities for ImMDI, with $\rho_{0}$ the saturation density, $E_0(\rho_{0})$ the binding energy at the saturation density, $K_0$ the incompressibility, $U_0^\infty$ the isoscalar potential in the nuclear matter at the saturation density and at infinitely large nucleon momentum, and $E_{sym}(\rho_{0})$ the symmetry energy at the saturation density.}
    \begin{tabular}{|c | c| c|c|}
   \hline
    $A_0$ (MeV)     & -66.963 & 92.144 & 100.466  \\
   \hline
    $B$ (MeV)     & 141.963 & 167.144 & 175.466  \\
   \hline
    $C_{u0}$ (MeV)    & -99.70 & -92.34 & -87.52 \\
   \hline
    $C_{l0}$ (MeV) & -60.49 & -52.34 & -47.19  \\
   \hline
    $\sigma$         & 1.2652 & 1.2646 & 1.2821\\
   \hline
    $\Lambda$ ($p_{f0}$)    & 2.424 & 3.401 & 5.369 \\
   \hline
   \hline
     $\rho_{0}$ (fm$^{-3}$) & 0.16 & 0.16 & 0.16 \\
   \hline
     $E_0(\rho_{0})$ (MeV) & -16 & -16 & -16 \\
   \hline
     $K_0$ (MeV)  & 230 & 230 & 230 \\
   \hline
     $U_0^\infty$ (MeV)   & 75 & 75 & 75 \\
   \hline
     $m_{s0}^*$ ($m$)  & 0.7 & 0.8 & 0.9  \\
   \hline
     $E_{sym}(\rho_{0})$ (MeV)   & 32.5 & 32.5 & 32.5 \\
   \hline
    \end{tabular}
  \label{T1}
\end{table}

\begin{figure}[ht]
	\includegraphics[scale=0.35]{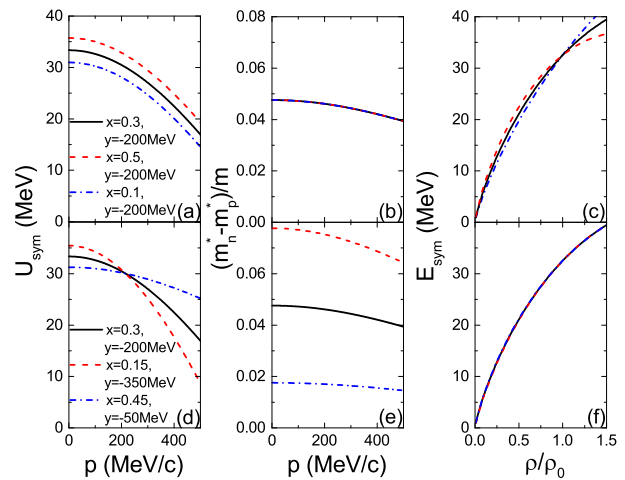}
	\caption{(Color online) Momentum dependence of the symmetry potential [(a), (d)] and the relative neutron-proton effective mass splitting [(b), (e)] in the isospin asymmetric nuclear matter at $\rho=0.1$ fm$^{-3}$ and $\delta=0.2$, as well as the density dependence of the symmetry energy [(c), (f)], from different parameter values of $x$ and $y$. } \label{fig2}
\end{figure}

Figure~\ref{fig2} displays the symmetry potential [$U_{sym}=(U_n-U_p)/2\delta$] and the relative neutron-proton effective mass splitting as a function of the nucleon momentum in the nuclear matter at $\rho=0.1$ fm$^{-3}$ and $\delta=0.2$, as well as the density dependence of the symmetry energy, by setting $m_{s0}=0.9m$ and other isoscalar parameters as listed in Table~\ref{T1}. Adjusting the $x$ parameter changes the momentum-independent part of the symmetry potential and the density dependence of the symmetry energy, while the neutron-proton effective mass splitting remains unaffected. It is seen that a larger symmetry energy at subsaturation densities corresponds to a stronger symmetry potential in this case. Adjusting the $y$ parameter alone changes both the momentum dependence of the symmetry potential and the density dependence of the symmetry energy~\cite{Xu15}. By adjusting both values of $x$ and $y$, it is possible to get very similar symmetry energies but different symmetry potentials and neutron-proton effective mass splittings. Again, since the low-momentum part dominates the dynamics in the simulation of nuclei resonances, a larger neutron-proton effective mass splitting generally leads to an overall stronger symmetry potential. The corresponding slope parameters $L$ of the symmetry energy and the isovector nucleon effective masses $m_{v0}^*$ from these $x$ and $y$ values are listed in Table~\ref{T2}.

\begin{table}\small
  \caption{Values of $x$ and $y$ parameters for ImMDI and the corresponding slope parameters $L$ of the symmetry energy and the isovector nucleon effective masses $m_{v0}^*$.}
    \begin{tabular}{|c | c| c|c|c|c|}
   \hline
    $x$      & 0.3 & 0.5 & 0.1 & 0.15 & 0.45 \\
   \hline
    $y$ (MeV)     & -200 & -200 & -200 & -350 & -50 \\
   \hline
   \hline
     $L$ (MeV) & 53 & 66 & 40 & 54 & 53 \\
   \hline
     $m_{v0}^*$ ($m$) & 0.79 & 0.79 & 0.79 & 0.73 & 0.86\\
   \hline
    \end{tabular}
  \label{T2}
\end{table}

Besides the bulk ImMDI interaction, we have also incorporated the density gradient interaction and the Coulomb interaction. The potential energy contribution of the density gradient interaction is
\begin{equation}
V^{\rm grad} = \frac{G_S}{2} (\nabla \rho)^2 - \frac{G_V}{2} [\nabla (\rho_n-\rho_p)]^2,
\end{equation}
where $G_S$ and $G_V$ are the isoscalar and the isovector density gradient coefficients, respectively. Although the Fock contribution of the finite-range term in the ImMDI interaction leads to the density-dependent density gradient coefficients in the density-matrix expansion framework~\cite{Xu10}, these coefficients are generally very different from the empirical values. In the present work we adopt $G_S=132$ MeV fm$^5$ and $G_V=5$ MeV fm$^5$ as in Ref.~\cite{MSL0}. The potential energy contribution of the Coulomb interaction is
\begin{equation}
V^{\rm coul}(\vec{r}) = \frac{e^2}{2} \int \frac{\rho_p(\vec{r})\rho_p(\vec{r}^\prime)}{|\vec{r}-\vec{r}^\prime|}d^3r^\prime - \frac{3}{4} e^2 \left[\frac{3\rho_p(\vec{r})}{\pi} \right]^{4/3},
\end{equation}
with the first term representing the direct contribution and the second term being the exchange contribution.

\subsection{An improved isospin-dependent Boltzmann-Uehling-Uhlenbeck transport approach}

\begin{widetext}
The IBUU transport model originating from Ref.~\cite{Bertsch88} basically solves numerically the isospin-dependent BUU equation
\begin{eqnarray}\label{buu}
&&\frac{\partial \tilde{f}_\tau(\vec{p}_1)}{\partial t} + \nabla_p U_\tau \cdot \nabla_r
\tilde{f}_\tau(\vec{p}_1) - \nabla_r U_\tau \cdot \nabla_p \tilde{f}_\tau(\vec{p}_1) = -(d-\frac{1}{2})
\int \frac{d^3p_2}{(2\pi)^3}
\frac{d^3p_1^\prime}{(2\pi)^3} \frac{d^3p_2^\prime}{(2\pi)^3}
\frac{d\sigma_{\tau,\tau}}{d\Omega} v_{rel} \notag \\
&\times&
[\tilde{f}_\tau(\vec{p}_1)\tilde{f}_\tau(\vec{p}_2)(1-\tilde{f}_\tau(\vec{p}_1^\prime))(1-\tilde{f}_\tau(\vec{p}_2^\prime))-
\tilde{f}_\tau(\vec{p}_1^\prime)\tilde{f}_\tau(\vec{p}_2^\prime)(1-\tilde{f}_\tau(\vec{p}_1))(1-\tilde{f}_\tau(\vec{p}_2))] \notag\\
&\times& (2\pi)^3
\delta^{(3)}(\vec{p}_1+\vec{p}_2-\vec{p}_1^\prime-\vec{p}_2^\prime)-d
\int \frac{d^3p_2}{(2\pi)^3} \frac{d^3p_1^\prime}{(2\pi)^3}
\frac{d^3p_2^\prime}{(2\pi)^3}
\frac{d\sigma_{\tau,-\tau}}{d\Omega} v_{rel}\notag \\
&\times&
[\tilde{f}_\tau(\vec{p}_1)\tilde{f}_{-\tau}(\vec{p}_2)(1-\tilde{f}_\tau(\vec{p}_1^\prime))(1-\tilde{f}_{-\tau}(\vec{p}_2^\prime))-
\tilde{f}_\tau(\vec{p}_1^\prime)\tilde{f}_{-\tau}(\vec{p}_2^\prime)(1-\tilde{f}_\tau(\vec{p}_1))(1-\tilde{f}_{-\tau}(\vec{p}_2))] \notag\\
&\times& (2\pi)^3
\delta^{(3)}(\vec{p}_1+\vec{p}_2-\vec{p}_1^\prime-\vec{p}_2^\prime).
\end{eqnarray}
In the above, $\tilde{f}$ is the occupation probability with $1-\tilde{f}$ representing the Pauli blocking effect,  $\frac{d\sigma}{d\Omega}$ is the nucleon-nucleon differential cross section, and $v_{rel}$ is the relative velocity of the two nucleons before the collision. The relation between the phase-space distribution function $f$ and the occupation probability $\tilde{f}$ is $f=d\tilde{f}$, with $d=2$ being the spin degeneracy.
\end{widetext}

The left-hand side of the above BUU equation describes the time evolution of the phase-space distribution function $f_\tau(\vec{r}, \vec{p})$ in the mean-field potential, and this can be approximately realized by solving the canonical equations of motion for test particles~\cite{Won82,Bertsch88}. In this approach, the phase-space distribution $f_\tau(\vec{r}, \vec{p})$ as well as the local density can be obtained by averaging $N_{TP}$ parallel collision events, i.e.,
\begin{eqnarray}
f_\tau(\vec{r}, \vec{p})&=&\frac{1}{N_{TP}}\sum_{i \in \tau}^{AN_{TP}} h(\vec{r}-\vec{r}_{i})\delta(\vec{p}-\vec{p}_{i}), \\
\rho_\tau (\vec{r})&=&\frac{1}{N_{TP}}\sum_{i \in \tau}^{AN_{TP}}h(\vec{r}-\vec{r}_{i}),
\end{eqnarray}
where $h$ is a smooth function in coordinate space, and $A$ is the number of real particles, with each represented by $N_{TP}$ test particles. The form of the smooth function $h$ is taken from that in the lattice Hamiltonian framework~\cite{Lenk89}, i.e., the phase-space distribution function $f_L$ and the density $\rho_L$ at the sites of a three-dimensional cubic lattice are expressed as
\begin{eqnarray}
f_{L,\tau}(\vec{r}_{\alpha},\vec{p}) &=& \sum_{i \in \tau}^{AN_{TP}}S(\vec{r}_{\alpha}-\vec{r}_i)\delta(\vec{p}-\vec{p}_{i}),\\
\rho_{L,\tau}(\vec{r}_{\alpha}) &=& \sum_{i \in \tau}^{AN_{TP}}S(\vec{r}_{\alpha}-\vec{r}_i).
\end{eqnarray}
In the above, $\alpha$ is the site index, $\vec{r}_{\alpha}$ is the position of the site $\alpha$, and $S$ is the shape function describing the contribution of a test particle at $\vec{r}_i$ to the value of the quantity at $\vec{r}_{\alpha}$, i.e.,
\begin{eqnarray}
S(\vec{r})=\frac{1}{N_{TP}(nl)^6}g(x)g(y)g(z)
\end{eqnarray}
with
\begin{eqnarray}
g(q)=(nl-|q|)\Theta(nl-|q|).
\end{eqnarray}
$l$ is the lattice spacing, $n$ determines the range of $S$, and $\Theta$ is the Heaviside function. We adopt the values of $l=1$ fm and $n=2$ in the present study.

After using the above smooth function for $f_{L,\tau}(\vec{r}_{\alpha},\vec{p})$ and $\rho_L(\vec{r}_{\alpha})$, the Hamiltonian of the system can be expressed as
\begin{equation}\label{htotal}
H=\sum_{i}^{AN_{TP}}\sqrt{\vec{p}_{i}^{2}+m^2}+N_{TP}\widetilde{V},
\end{equation}
with the total potential energy expressed as
\begin{equation}
\widetilde{V}=l^3\sum_{\alpha}(V^{\rm ImMDI}_{\alpha} + V^{\rm grad}_{\alpha} + V^{\rm coul}_{\alpha}),
\end{equation}
where
\begin{eqnarray}
V^{\rm ImMDI}_{\alpha} &=& \frac{A_{u}\rho _{L,n}(\vec{r}_{\alpha})\rho _{L,p}(\vec{r}_{\alpha})}{\rho _{0}}+\frac{A_{l}}{2\rho _{0}}[\rho _{L,n}^{2}(\vec{r}_{\alpha})             \notag \\
& & +\rho _{L,p}^{2}(\vec{r}_{\alpha})]+\frac{B}{\sigma+1}\frac{\rho_{L}^{\sigma +1}(\vec{r}_{\alpha})}{\rho _{0}^{\sigma }}[1-x\delta_L ^{2}(\vec{r}_{\alpha})]+\frac{1}{\rho _{0}}                       \notag \\
& & \times \sum_{i,j} \sum_{\tau_{i} ,\tau_{j}}C_{\tau_{i} ,\tau_{j}}
	\frac{S(\vec{r}_{\alpha}-\vec{r}_i)S(\vec{r}_{\alpha}-\vec{r}_j)}{1+(\vec{p}_{i}
	-\vec{p}_{j})^{2}/\Lambda ^{2}} , \label{vimmdi}
\end{eqnarray}
\begin{equation}
V^{\rm grad}_{\alpha} = \frac{G_S}{2} [\nabla \rho_L(\vec{r}_{\alpha})]^2 - \frac{G_V}{2} \{\nabla [\rho_{L,n}(\vec{r}_{\alpha})-\rho_{L,p}(\vec{r}_{\alpha})]\}^2,\label{vgrad}
\end{equation}
\begin{eqnarray}
V^{\rm coul}_{\alpha} &=& \frac{e^2}{2}l^3 \sum_{\alpha^\prime} \frac{\rho_{L,p}(\vec{r}_{\alpha})\rho_{L,p}(\vec{r}_{\alpha^\prime})}{|\vec{r}_\alpha - \vec{r}_{\alpha^\prime}|} - \frac{3}{4} e^2 \left[\frac{3\rho_{L,p}(\vec{r}_\alpha)}{\pi} \right]^{4/3}\notag\\
&-& \frac{e^2}{2}l^3 \sum_{\alpha^\prime}  \sum_{i \in p}\frac{S(\vec{r}_{\alpha}-\vec{r}_i)S(\vec{r}_{\alpha^\prime}-\vec{r}_i)}{|\vec{r}_\alpha - \vec{r}_{\alpha^\prime}|}\label{vcoul}
\end{eqnarray}
are the corresponding contributions of the ImMDI interaction, the density gradient interaction, and the Coulomb interaction, respectively. $\delta_L(\vec{r}_{\alpha})=[\rho _{L,n}(\vec{r}_{\alpha})-\rho _{L,p}(\vec{r}_{\alpha})]/[\rho _{L,n}(\vec{r}_{\alpha})+\rho _{L,p}(\vec{r}_{\alpha})]$ is the isospin asymmetry at $\vec{r}_{\alpha}$ with $\rho _{L,n}(\vec{r}_{\alpha})$ and $\rho _{L,p}(\vec{r}_{\alpha})$ being respectively the number density of neutrons and protons there, and the third term in Eq.~(\ref{vcoul}) subtracts the self contribution of the Coulomb interaction from the same proton due to its finite size in the lattice Hamiltonian framework. The canonical equations of motion for the $i$th test particle from the above Hamiltonian can thus be written as
\begin{eqnarray}
 \frac{d\vec{r}_{i}}{dt}&=&\frac{\partial H}{\partial\vec{p}_{i}}
 = \frac{\vec{p}_i}{\sqrt{\vec{p}_{i}^{2}+m^2}} + N_{TP}\frac{\partial\widetilde{V}}{\partial\vec{p}_{i}},  \label{rt}\\
 \frac{d\vec{p}_{i}}{dt} &=&-\frac{\partial H}{\partial\vec{r}_{i}}
 = -N_{TP}\frac{\partial\widetilde{V}}{\partial\vec{r}_{i}}.  \label{pt}
\end{eqnarray}

Further improvements have been incorporated into the IBUU transport approach. The coordinates of initial neutrons and protons are sampled uniformly within a sphere of the radius $R_n$ and $R_p$ respectively. The initial momenta are sampled within the local isospin-dependent Fermi sphere. The values of $R_n$ and $R_p$ are adjusted to reproduce the minimum total energy of the system calculated according to Eq.~(\ref{htotal}), so that the ground state of the system can be achieved as in Ref.~\cite{Lenk89}. In addition, a special treatment is applied in nucleon-nucleon collisions in order to guarantee that the energy conservation condition is satisfied in each collision within numerical errors even with the momentum-dependent potential, and this is detailed in Appendix~\ref{econ}. We have also improved the Pauli blocking treatment by calculating the isospin-dependent occupation probability in the local frame rather than in the collisional frame, and this, together with the previous interpolation method, helps to enhance the Pauli blocking rate.

\subsection{Nuclei giant resonances}

In the present study, we mainly focus on the ISGQR and the IVGDR in $^{208}$Pb. Their corresponding operators can be written respectively as
\begin{eqnarray}
\hat{Q}_{\rm ISGQR} &=& \frac{1}{A} \sum_{i=1}^A \sqrt{\frac{5}{16\pi}}(2\hat{z}_i^2-\hat{x}_i^2-\hat{y}_i^2), \label{QISGQR}\\
\hat{Q}_{\rm IVGDR} &=& \frac{N}{A} \sum_{i=1}^Z \hat{z}_i - \frac{Z}{A} \sum_{i=1}^N \hat{z}_i, \label{QIVGDR}
\end{eqnarray}
where $N$, $Z$, and $A$ are respectively the neutron, proton, and nucleon numbers in a nucleus. In the linear response region, the oscillation frequency of the nucleus resonance is independent of the way the nucleus is initially excited. For the ISGQR, nucleons in the nucleus are initially excited as
\begin{eqnarray}
&x_i \rightarrow x_i/\lambda, y_i \rightarrow y_i/\lambda, z_i \rightarrow z_i\lambda^2,\label{isgqrini1}\\
&(p_{x})_i \rightarrow (p_x)_i\lambda, (p_y)_i \rightarrow (p_y)_i\lambda, (p_z)_i \rightarrow (p_z)_i/\lambda^2,\label{isgqrini2}
\end{eqnarray}
where $\lambda=1.1$ is the small perturbation parameter. For the IVGDR, we adopt the standard way of the initial excitation~\cite{Urb12}
\begin{eqnarray}
\vec{r}_i &\rightarrow& \vec{r}_i + \eta\frac{\partial q(\vec{r}_i,\vec{p}_i)}{\partial \vec{p}_i}, \label{ivgdrini1}\\
\vec{p}_i &\rightarrow& \vec{p}_i - \eta\frac{\partial q(\vec{r}_i,\vec{p}_i)}{\partial \vec{r}_i}, \label{ivgdrini2}
\end{eqnarray}
where $\eta=25$ MeV/c is the small perturbation constant, and
\begin{eqnarray}
q^{}_{\rm IVGDR}(\vec{r}_i,\vec{p}_i) = \left\{\begin{matrix}
\frac{N}{A}z_i   ~~\rm (protons)\\
-\frac{Z}{A}z_i   ~~\rm (neutrons)\label{ivgdrini3}
\end{matrix}\right.,
\end{eqnarray}
can be obtained from Eq.~(\ref{QIVGDR}).

With the time evolution of the corresponding moment $Q(t)$ from IBUU transport simulations, the strength function of the IVGDR can be obtained from
\begin{equation}\label{se}
S(E) = -\frac{1}{\pi\eta} \int_0^\infty dt Q(t) \sin(Et).
\end{equation}
By calculating the moments of the strength function
\begin{equation}
m_k = \int_0^\infty dE E^k S(E),
\end{equation}
one can compare the transport simulation results with the available experimental data. For example, the centroid energy $E_{-1}$ and the electric dipole polarizability $\alpha_D$ can be obtained respectively from
\begin{eqnarray}
E_{-1} &=& \sqrt{m_1/m_{-1}},\\
\alpha_D &=& 2e^2m_{-1}.
\end{eqnarray}

\section{Results and discussions}
\label{results}

In the present study, we reproduce both the excitation energy and the decay width of the ISGQR in $^{208}$Pb measured experimentally, by adjusting the isoscalar nucleon effective mass $m_{s0}^*$ and a constant and isotropic nucleon-nucleon scattering cross section. Using the same $m_{s0}^*$ and the cross section, we further constrained the symmetry energy and the neutron-proton effective mass splitting using the centroid energy and the electric dipole polarizability extracted from the IVGDR in $^{208}$Pb, by comparing results from the IBUU transport approach with the experimental data. We use several IBUU runs for each scenario, and 200 test particles are used for each run. The statistical errors are calculated based on results of different IBUU runs. The moments of the ISGQR and the IVGDR are calculated from binded nucleons with their local densities higher than $\rho_0/20$.

\subsection{Isoscalar giant quadruple resonance}

\begin{figure}[ht]
	\includegraphics[scale=0.3]{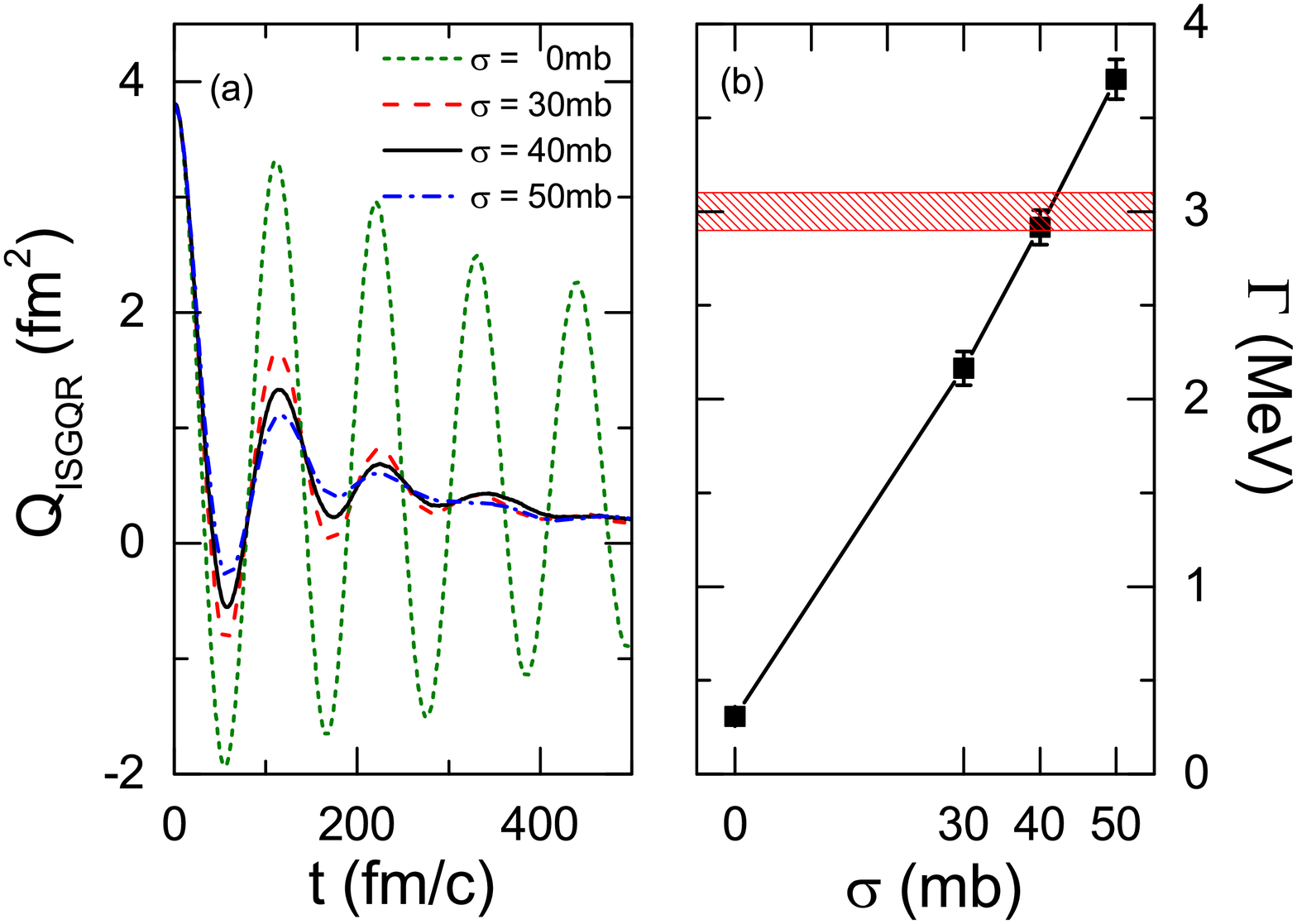}
	\caption{(Color online) Time evolution of the ISGQR moment (a) and the decay width of the ISGQR (b) from different nucleon-nucleon cross sections. The experimentally measured width~\cite{ISGQRexnew} is plotted as a band for comparison.} \label{fig3}
\end{figure}

With the initial $^{208}$Pb nucleus excited according to Eqs.~(\ref{isgqrini1}) and (\ref{isgqrini2}), the time evolutions of the ISGQR moment using different nucleon-nucleon cross sections are compared in Fig.~\ref{fig3}(a), by using the parameters with $m_{s0}^*=0.9m$ as listed in Table~\ref{T1}. It is obviously seen that a larger nucleon-nucleon cross section leads to a stronger damping of the ISGQR oscillation, since more attempted and successful nucleon-nucleon collisions occur. Even in the Vlasov calculation with $\sigma=0$ mb, the oscillation mode damps very slowly due to the Landau damping mechanism. On the other hand, the oscillation frequency is seen to be not much affected by the nucleon-nucleon cross section. It is interesting to see that the moment does not return to zero especially with larger cross sections. From the observation, the ISGQR moment generally shows a periodical oscillation behavior with an exponential decay, so it can be fitted with the following function~\cite{Gai10}
\begin{equation}\label{isgqrfit}
Q_{\rm ISGQR}(t) = a\sin[b(t-t_0)]\exp(-ct)+d,
\end{equation}
where $a$ represents the oscillation magnitude, $b$ represents the oscillation frequency, $t_0$ represents the initial oscillation phase, $c$ represents the decay width, and $d$ represents some possible average displacement. The resulting decay widths $\Gamma \sim c$ for different nucleon-nucleon cross sections are shown in Fig.~\ref{fig3}(b). The larger decay width from the larger cross section is intuitively understandable. Even in the Vlasov scenario, the decay width is non-zero. In the present study, we invoke the experimental results of the ISGQR extracted in Ref.~\cite{ISGQRexnew}, where the decay width is $3.0 \pm 0.1$ MeV shown as the band in Fig.~\ref{fig3}(b). The cross section $\sigma=40$ mb reproduces this decay width reasonably well, and the collision effect is seen to be similar to that from the particle-vibration coupling~\cite{Roc13a}.

\begin{figure}[ht]
	\includegraphics[scale=0.3]{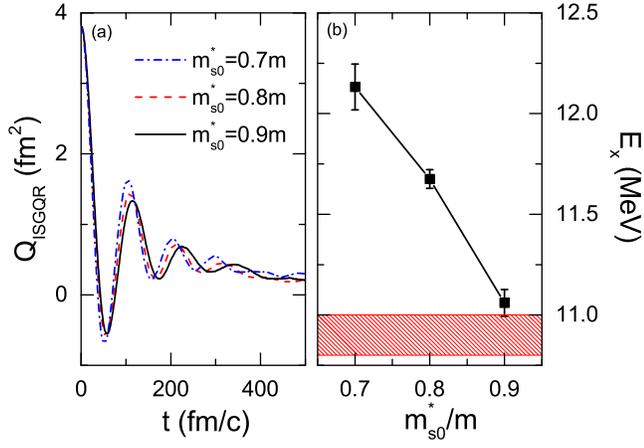}
	\caption{(Color online) Time evolution of the ISGQR moment (a) and the excitation energy of the ISGQR (b) from different isoscalar nucleon effective masses. The experimentally measured excitation energy~\cite{ISGQRexnew} is plotted as a band for comparison.} \label{fig4}
\end{figure}

Using different isoscalar nucleon effective masses $m_{s0}^*$, the time evolutions of the ISGQR moment are compared in Fig.~\ref{fig4}(a), where the nucleon-nucleon cross section $\sigma=40$ mb is used in each scenario. The different oscillation frequencies from different $m_{s0}^*$ can already be seen from the time evolution of the ISGQR moment. Fitting the ISGQR moment with Eq.~(\ref{isgqrfit}), the excitation energies $E_x \sim b$ from different $m_{s0}^*$ are shown in Fig.~\ref{fig4}(b). It is seen that a larger $m_{s0}^*$ leads to a smaller $E_x$. This is understandable from Fig.~\ref{fig1}, since a smaller $m_{s0}^*$ leads to a more attractive isoscalar potential below the Fermi momentum, serving as a stronger restoring force of the ISGQR and increasing the oscillation frequency. The experimental measured excitation energy $E_x=10.9 \pm 0.1$~\cite{ISGQRexnew} is represented by the band in Fig.~\ref{fig4}(b), which is reproduced reasonably well with the parameterization $m_{s0}^*=0.9m$.

\subsection{Isovector giant dipole resonance}

\begin{figure}[ht]
	\includegraphics[scale=0.3]{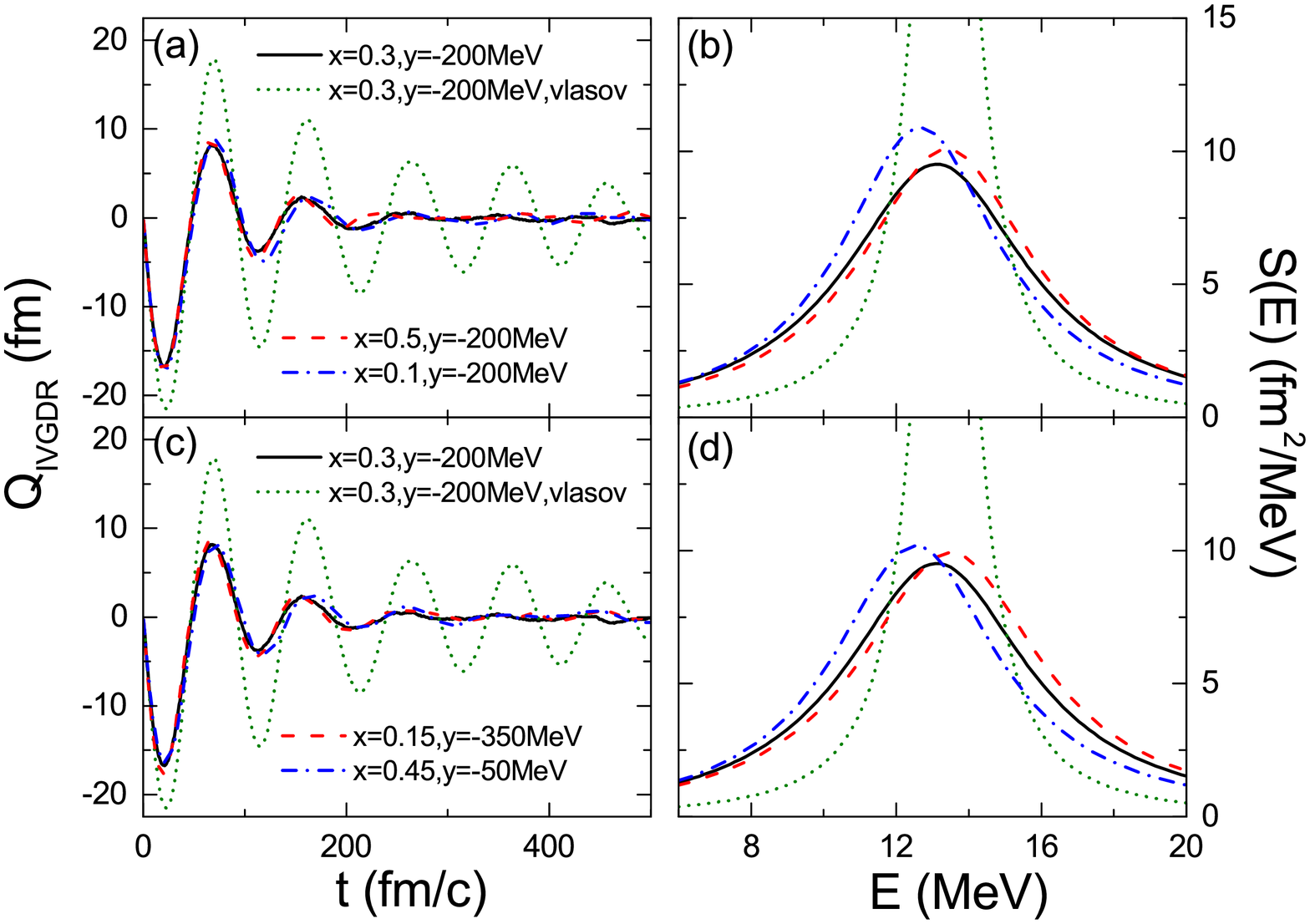}
	\caption{(Color online) Time evolution of the IVGDR moment (left) and the strength function of the IVGDR (right) from different scenarios.} \label{fig5}
\end{figure}

Using the same nucleon-nucleon cross section $\sigma=40$ mb and the initial excitation as Eqs.~(\ref{ivgdrini1}), (\ref{ivgdrini2}), and (\ref{ivgdrini3}), we have stimulated the IVGDR in $^{208}$Pb, and the time evolutions of the moment from different scenarios are displayed in Fig.~\ref{fig5}(a) and Fig.~\ref{fig5}(c). The periodic oscillation and decay behavior of the IVGDR moment in all scenarios can be fitted with the following form
\begin{equation}
Q_{\rm IVGDR}(t) = a\sin(bt)\exp(-ct).
\end{equation}
The advantage of the fitting is that the same oscillation behavior is extrapolated to infinity time and the integral in Eq.~(\ref{se}) can be carried out analytically, i.e.,
\begin{equation}\label{sivgdr}
S(E)=\frac{ac}{2\pi \eta }\left [ \frac{1}{c^{2}+( b+E)^{2}} - \frac{1}{c^{2}+( b-E)^{2}} \right].
\end{equation}
The resulting strength functions from different scenarios are shown in Fig.~\ref{fig5}(b) and Fig.~\ref{fig5}(d). The different $x$ and $y$ values corresponds to different symmetry energies and neutron-proton effective mass splittings, essentially different symmetry potentials, as shown in Fig.~\ref{fig2} and Table~\ref{T2}. Results with different symmetry energies but the same neutron-proton effective mass splitting are thus compared in the upper panels of Fig.~\ref{fig5}, while results with the same symmetry energy but different neutron-proton effective mass splittings are compared in the lower panel of Fig.~\ref{fig5}. The effects can all be understood from the low-momentum part of the symmetry potential, which is the dominating restoring force of the IVGDR. Since the cases ($x=0.5,y=-200$MeV) and ($x=0.15,y=-350$MeV) have a stronger symmetry potential at low momenta compared respectively with ($x=0.1,y=-200$MeV) and ($x=0.45,y=-50$MeV), the former ones have a higher peak frequency of the IVGDR, as shown in their strength functions. Again, the Vlasov mode without nucleon-nucleon collisions shows an oscillation with a larger magnitude and a weak damping, leading to a very sharp strength function with a small width. The shape of the strength function extracted experimentally~\cite{Tam11} is similar to those with nucleon-nucleon collisions rather than that from the Vlasov mode.

\begin{figure}[ht]
	\includegraphics[scale=0.3]{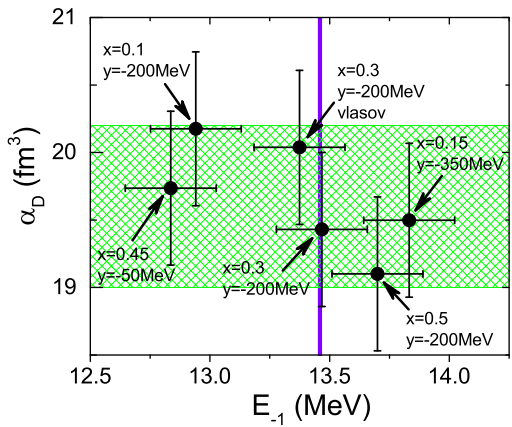}
	\caption{(Color online) The resulting centroid energies $E_{-1}$ and the electric dipole polarizability $\alpha_D$ from different scenarios compared with the experimental results~\cite{IVGDRe,Tam11,Roc15} shown as bands.} \label{fig6}
\end{figure}

With the analytical formula of the strength function Eq.~(\ref{sivgdr}), the moments as well as other observables can also be expressed analytically as
\begin{eqnarray}
m_{-1}&=&\frac{-ab}{2\eta\left ( b^{2}+c^{2} \right )},  \\
m_{1}&=&\frac{-ab}{2\eta}, \\
E_{-1}&=& \sqrt{b^{2}+c^{2}}, \\
\alpha_{D}&=&\frac{-e^{2}ab}{\eta\left ( b^{2}+c^{2} \right )}.
\end{eqnarray}
Figure~\ref{fig6} displays the resulting centroid energies $E_{-1}$ and the electric dipole polarizabilities $\alpha_D$ for the corresponding scenarios as in Fig.~\ref{fig5}. The experimental results of $E_{-1}=13.46$ MeV from photoabsorption reactions~\cite{IVGDRe}, and $\alpha_D=19.6 \pm 0.6$ fm$^3$, which is measured from photoabsorption cross sections as well as polarized proton inelastic scatterings~\cite{Tam11} and further corrected by subtracting the contribution of quasideuteron excitations~\cite{Roc15}, are also plotted for comparison. It is seen that the electric dipole polarizability can generally be reproduced with the parameterization adopted here, while the centroid energy gives a very stringent constraints on the $x$ and $y$ parameters. Comparing the results with and without nucleon-nucleon collisions, it is seen that the centroid energies are very similar within statistical errors, while a considerable effect on the electric dipole polarizability is observed, as a result of the different shapes of the strength function shown in Fig.~\ref{fig5}.

\begin{figure}[ht]
	\includegraphics[scale=0.3]{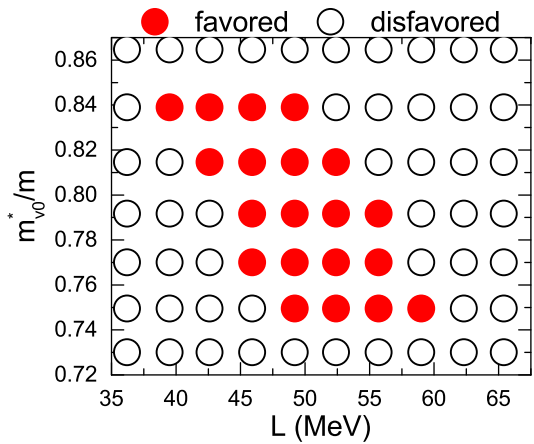}
	\caption{(Color online) Favored and disfavored values of the slope parameter $L$ of the symmetry energy and the isovector nucleon effective mass $m_{v0}^*$ from the experimental data of $E_{-1}$ and $\alpha_D$.} \label{fig7}
\end{figure}

The favored $x$ and $y$ values can be obtained by comparing the resulting $E_{-1}$ and $\alpha_D$ with the experimental data, in the way as shown in Fig.~\ref{fig6}. Using the same isoscalar parameterization with $m_{s0}^*=0.9m$ as shown in Table~\ref{T1}, the favored $x$ and $y$ values can be mapped in the two-dimensional plane of the slope parameter $L$ of the symmetry energy and the isovector nucleon effective mass $m_{v0}^*$, as displayed in Fig.~\ref{fig7}. The anticorrelation relation between $L$ and $m_{v0}^*$ is observed.
It is seen that the favored values of $L$ and $m_{v0}^*$ are within an area of about $36<L<62$ MeV and $0.73<m_{v0}^*/m<0.86$. The later corresponds to the range of the neutron-proton effective mass splitting $0.08\delta<(m_{n0}^*-m_{p0}^*)/m<0.42\delta$ at the saturation density and at the Fermi momentum. The constraint on $L$ further narrows down the recent constraint of $L=58.7 \pm 28.1$ MeV~\cite{BAL13,Oer17}, and the constraint on the neutron-proton effective mass splitting is consistent with those obtained from various approaches in the literature~\cite{BAL13,Li15,Hol16,zhangzhen16,Kon17,Bal17,Mon17}.

\section{summary}
\label{summary}

Based on an improved isospin-dependent Boltzmann-Uehling-Uhlenbeck transport approach and using an improved isospin- and momentum-dependent interaction, we have studied the isoscalar giant quadrupole resonance (ISGQR) and the isovector giant dipole resonance (IVGDR) in $^{208}$Pb. The width of the strength function and the excitation energy of the ISGQR are reproduced respectively by choosing a proper nucleon-nucleon cross section $\sigma=40$ mb and isoscalar nucleon effective mass $m_{s0}^*=0.9m$. With the same $\sigma$ and $m_{s0}^0$, we have further constrained the slope parameter $L$ of the symmetry energy and the isovector nucleon effective mass $m_{v0}^*$, by comparing the resulting centroid energy and the electric dipole polarizability, extracted from the strength function of the IVGDR, with the corresponding experimental data. The isoscalar potential and the symmetry potential below the Fermi momentum dominate the restoring force of the ISGQR and IVGDR. Incorporating the nucleon-nucleon collisions leads to almost the same peak energy of the strength function but broads it by damping the collective oscillation of the IVGDR, and thus has considerable effects on the resulting electric dipole polarizability. The favored values of $L$ and $m_{v0}^*$ are within an area of about $36<L<62$ MeV and $0.73<m_{v0}^*/m<0.86$, where they are anticorrelated with each other. The latter leads to the neutron-proton effective mass splitting $0.08\delta<(m_{n0}^*-m_{p0}^*)/m<0.42\delta$. Although the experimental measured centroid energy of the IVGDR gives a stringent constraint on $L$ and $m_{v0}^*$, further efforts on measuring more accurately the electric dipole polarizability is encouraged to pin down nuclear interactions in the isovector channel.

\begin{acknowledgments}

This work was supported by the National Natural Science
Foundation of China under Grant No. 11922514. Helpful discussions with Rui Wang are acknowledged.

\end{acknowledgments}

\begin{appendix}

\section{Energy conservation in nucleon-nucleon collisions with a momentum-dependent potential}
\label{econ}

Although the Bertsch's prescription~\cite{Bertsch88} conserves the energy in each nucleon-nucleon (NN) collision in free space, this is not the case in the presence of the momentum-dependent potential. This is because the contribution of the momentum-dependent part in Eq.~(\ref{vimmdi}) generally changes after a NN collision, due to their different final nucleon momenta compared to those before the NN collision. As a remedy, we modified the Bertsch's prescription in the following way.

The collision between nucleon 1 and nucleon 2 happens in their center-of-mass (C.M.) frame. In the original Bertsch's prescription, the momentum in the C.M. frame changes its direction while keeping its magnitude after a successful NN collision, and their final momenta $\vec{p}_{1,2}$ and kinetic energies $E_{1,2}=\sqrt{\vec{p}_{1,2}^2+m^2}$ are from the Lorentz transformation back to the collisional frame according to
\begin{eqnarray}
\vec{p}_{1,2} &=& \gamma (\pm \vec{p}^{}_{\rm CM} + \vec{\beta} E_{\rm CM}),\\
E_{1,2} &=& \gamma (E_{\rm CM} \mp \vec{\beta} \cdot \vec{p}^{}_{\rm CM}).
\end{eqnarray}
In the above, $E_{\rm CM}=\sqrt{\vec{p}_{\rm CM}^2+m^2}$ is the kinetic energy in the C.M. frame with $\vec{p}^{}_{\rm CM}$ being the momentum after the collision, $\gamma=1/\sqrt{1-\beta^2}$ is the Lorentz factor with $\vec{\beta}$ being the velocity of the C.M. frame with respect to the collisional frame. The upper (lower) signs in the above equations are for nucleon 1(2). As mentioned before, this prescription conserves the total momentum and kinetic energy. In order to conserve both the total momentum and total energy in the presence of the momentum-dependent potential, we modify the prescription by changing the magnitudes of $\vec{p}^{}_{\rm CM}$ and $\vec{\beta}$ while keeping their direction, i.e., $\vec{p}^{\prime}_{\rm CM}=c_1\vec{p}^{}_{\rm CM}$ and $\vec{\beta}^\prime = c_2\vec{\beta}$ where $c_1$ and $c_2$ are constants to be determined, and the Lorentz transformation from the C.M. frame back to the collisional frame is now expressed as
\begin{eqnarray}
\vec{p}^\prime_{1,2} &=& \gamma^\prime (\pm \vec{p}^{\prime}_{\rm CM} + \vec{\beta}^\prime E^\prime_{\rm CM}),\\
E^\prime_{1,2} &=& \gamma^\prime (E^\prime_{\rm CM} \mp \vec{\beta}^\prime \cdot \vec{p}^{\prime}_{\rm CM}),
\end{eqnarray}
with $\gamma^\prime=1/\sqrt{1-{\beta^\prime}^2}$. To satisfy the momentum and energy conservation conditions, we need to solve the following equations
\begin{eqnarray}
\vec{p}_1 + \vec{p}_2 &=& \vec{p}^\prime_1 + \vec{p}^\prime_2,  \label{pcons}\\
E_1 + E_2 + v_{12} &=& E^\prime_1 + E^\prime_2 + v^\prime_{12}, \label{econs}
\end{eqnarray}
where
\begin{eqnarray}
v^{(\prime)}_{12} &=& \frac{2l^3N_{TP}}{\rho _{0}}\left[ \sum_{j} \sum_{\tau_{j}}C_{\tau_{1} ,\tau_{j}}
	\frac{S(\vec{r}_{\alpha}-\vec{r}_1)S(\vec{r}_{\alpha}-\vec{r}_j)}{1+(\vec{p}^{(\prime)}_{1}
	-\vec{p}_{j})^{2}/\Lambda ^{2}} \right. \notag\\
  &+& \left. \sum_{i} \sum_{\tau_{i}}C_{\tau_{i} ,\tau_{2}}
	\frac{S(\vec{r}_{\alpha}-\vec{r}_i)S(\vec{r}_{\alpha}-\vec{r}_2)}{1+(\vec{p}_{i}
	-\vec{p}^{(\prime)}_{2})^{2}/\Lambda ^{2}} \right]
\end{eqnarray}
is the contribution of nucleon 1 and nucleon 2 to the momentum-dependent part of the potential energy in Eq.~(\ref{vimmdi}). Equations~(\ref{pcons}) and (\ref{econs}) can be solved numerically using the iteration method by starting from $c_1=c_2=1$.
The above method guarantees the momentum and energy conservation in each NN collision in the presence of the momentum-dependent potential, and can be easily generalized to the case of inelastic collisions.
\end{appendix}

\end{document}